\documentclass[twocolumn,showpacs,prl]{revtex4}

\usepackage{amsmath,amssymb}
\usepackage{graphicx}
\usepackage{verbatim}

\draft
\begin{document}

\title{Fractional Spin Hall Effect in Atomic Bose Gases}

\author{Xiong-Jun Liu$^{a,b}$\footnote{Electronic address:
phylx@physics.tamu.edu}, Xin Liu$^{b}$, L. C. Kwek$^{a,c}$ and C.
H. Oh$^{a}$\footnote{Electronic address: phyohch@nus.edu.sg}}
\affiliation{a. Department of Physics, National University of
Singapore,
2 Science Drive 3, Singapore 117542 \\
b. Department of Physics, Texas A\&M University, College Station,
Texas 77843-4242, USA\\
c. National Institute of Education, Nanyang Technological
University, Singapore 639798}

\begin{abstract}
We propose fractional spin hall effect (FSHE) by coupling
pseudospin states of cold bosonic atoms to optical fields. The
present scheme is an extension to interacting bosonic system of
the recent work \cite{liu,zhu} on optically induced spin hall
effect in non-interacting atomic system. The system has two
different types of ground states. The 1st type of ground state is
a $1/3$-factor Laughlin function, and has the property of chiral -
anti-chiral interchange antisymmetry, while the 2nd type is shown
to be a $1/4$-factor wave function with chiral - anti-chiral
symmetry. The fractional statistics corresponding to the
fractional spin Hall states are studied in detail, and are
discovered to be different from that corresponding to the
fractional quantum Hall (FQH) states. Therefore the present FSHE
can be distinguished from FQH regime in the measurement.
\end{abstract}
\date{\today }
\maketitle

\indent

\section{Introduction}

Intrinsic spin Hall effect (SHE) has attracted great attention
since it was predicted in semiconductors with spin-orbit coupled
structures \cite{SHE1,SHE2,SHE3,graphene}, with the concomitant
creation of spin currents and realization of quantized spin hall
conductance (SHC). Quantum SHE with non-interacting particles was
firstly studied in graphene \cite{graphene1,graphene2} and
semiconductors with a strain gradient structure \cite{zhang2},
while by now there are no experimental systems available for such
proposals. Recently, Bernevig, Hughes and Zhang theoretically
predicted the quantum SHE in HgTe/CdTe quantum wells
\cite{zhang1}. By varying the thickness of the quantum well, a
quantum phase transition is obtained between the conventional
insulator and the quantum spin Hall (QSH) insulator. Such a
prediction has been remarkably confirmed in the recent experiment
\cite{experiment1}. The QSH insulator is a topologically
nontrivial state of matter protected by the time reversal
symmetry, and it is currently described through a $Z_2$
classification \cite{graphene1,graphene2,Z2}. Considering the
nontrivial topological properties, such QSH insulators may have
not only potential applications and but also the fundamental
importance in physics.

On the other hand, the similar idea for the SHE has been proposed
in cold non-interacting atomic system by coupling the internal
atomic states (atomic spins) to radiation \cite{liu,zhu}. The
atom-light coupling creates a spin-dependent effective magnetic
field, leading to SHE in fermionic atomic systems. A challenging
but interesting extension is the realization of fractional spin
Hall (FSH) regime with the particle-particle interactions
considered. The correlated many-body function in the FSH regime
was initially described in the Ref. \cite{zhang2}. Nevertheless,
many issues are left in the fractional spin Hall effect (FSHE),
e.g. the fractional statistics corresponding to the FSH state is
not clear and needs to be further investigated. Comparing with
solid matters, ultra-cold atomic system provides a unique access
to the study of complex many-body dynamics with its extremely
clean environment and remarkable controllability in the
parameters. Therefore it is very suggestive to study the FSHE by
extending optically induced SHE \cite{liu,zhu} to interacting
bosonic atomic systems where, different from former schemes with
the non-interacting atomic gas, the nonlinear interaction between
atoms (s-wave scattering) plays a central role in the Hall effect.

In this paper, we propose FSHE by coupling internal electronic
states of cold bosonic atoms to the external optical fields, with
atom-atom interaction considered. Under the lowest Landau level
(LLL) condition, we can exactly study the ground states of the
present many-body system. The intriguing fundamental properties of
FSH states and the corresponding fractional statistics in our
system are investigated.

The paper is organized as follow. In section II, we derive the
effective Hamiltonian that gives FSHE. Then in section III, we
study the FSH state and corresponding quasi-particle excitation,
with which we point out differences between the present FSH regime
and the fractional quantum Hall (FQH) regime. Realization of the
FSHE in realistic atomic systems is discussed in section IV.
Finally we conclude our results in section V.

\section{Effective Hamiltonian}

In this section we shall study two different configurations to
obtain the effective Hamiltonian that gives the FSHE in the cold
atoms.

\subsection{Four-level configuration}

We first consider the four-level configuration shown in Fig. 1(a).
An ensemble of cold bosonic atoms with four internal angular
momentum states (atomic spins), described by atomic state
functions $\Phi_\alpha(\bold r,t)$ ($\alpha=e_\pm,s_\pm$),
interact with two external light fields.
\begin{figure}[ht]
\includegraphics[width=0.9\columnwidth]{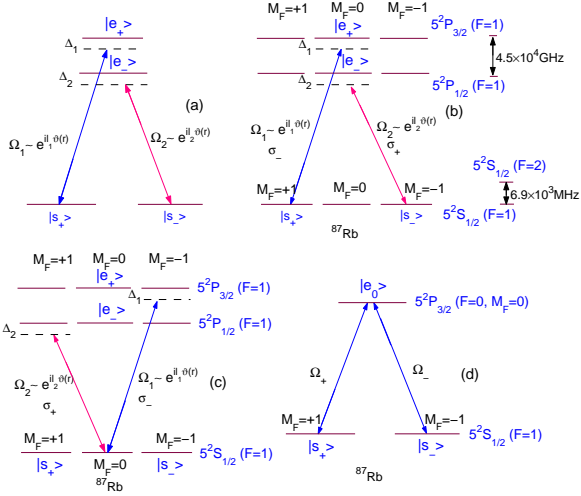}
\caption{(color online) (a) Four-level bosonic atoms interacting
with two light fields; This case can be experimentally realized
with e.g. $^{87}$Rb atoms ((b) and (c) ); (d) Initial condition
achieved by pumping the atoms into $|S_\pm\rangle$ with
$|\Omega_+|=|\Omega_-|$ through the $\Lambda$-type configuration.}
\label{fig1}\end{figure}
The transitions from $|s_\pm\rangle$ to
$|e_\pm\rangle$ are respectively coupled by a $\sigma_-$ light
with the Rabi-frequency $\Omega_1=\Omega_{10}\exp(i(\bold
k_1\cdot\bold r+l_1\vartheta))$ and by a $\sigma_+$ light with the
Rabi-frequency $\Omega_2=\Omega_{20}\exp(i(\bold k_2\cdot\bold
r+l_2\vartheta))$, where $\bold k_{1,2}=k_{1,2}\bold{\hat e}_z$
and $\vartheta=\tan^{-1}({y}/{x})$. $l_1$ and $l_2$ indicate that
$\sigma_+$ and $\sigma_-$ photons respectively have the orbital
angular momenta $\hbar l_1$ and $\hbar l_2$ along the $+z$
direction \cite{angular}. It is convenient to introduce the
slowly-varying amplitudes of atomic wave-functions by (note
$\omega_{s_\pm}=0$): $\phi_{s_\pm}=\Phi_{s_\pm},
\phi_{e_+}=\Phi_{e_+}(\bold r,t)e^{-i(\bold k_1\cdot\bold
r-(\omega_{e_+}-\Delta_1)t)}, \phi_{e_-}=\Phi_{e_-}(\bold
r,t)e^{-i(\bold k_2\cdot\bold r-(\omega_{e_-}-\Delta_2)t)}$, where
$\hbar\omega_\alpha$ is the energy of the state $|\alpha\rangle$,
$\Delta_1$ and $\Delta_2$ are transition detunings. The total
Hamiltonian of the present system can be written as
$H=H_0+H_1+H_2$, with
\begin{eqnarray}\label{eqn:Hamiltonian1}
H_0&=&\sum_{\alpha=e_{\pm},s_{\pm}}\int
d^3r\phi_{\alpha}^*\bigl(-\frac{\hbar^2}{2m}\nabla^2+V(\bold
r)\bigl)
\phi_{\alpha}\nonumber\\
&+&\sum_{\alpha,\beta}\int d^3rd^3r'\phi_\alpha^*(\bold
r)\phi_\beta^*(\bold r')U_{\alpha\beta}(\bold r,\bold r')
\phi_\alpha(\bold r)\phi_\beta(\bold r'),\nonumber\\
H_1&=&\hbar\Delta_1\int d^3r\phi^*_{e_+}S_{e_+e_+}\phi_{e_+}\nonumber\\
&&-\hbar\int
d^3r(\phi^*_{e_+}\Omega_{10}e^{il_1\vartheta}S_{1+}\phi_{s_+}+h.a.),
\end{eqnarray}
\begin{eqnarray}\label{eqn:Hamiltonian01}
H_2&=&\hbar\Delta_2\int d^3r\phi^*_{e_-}S_{e_-e_-}\phi_{e_-}\nonumber\\
&&-\hbar\int
d^3r(\phi^*_{e_-}\Omega_{20}e^{il_2\vartheta}S_{2+}\phi_{s_-}+h.a.),\nonumber
\end{eqnarray}
with the atomic operators defined by
$S_{e_{\pm}e_{\pm}}=|e_{\pm}\rangle\langle e_{\pm}|$,
$S_{1+}=|e_+\rangle\langle s_+|, S_{2+}=|e_-\rangle\langle s_-|$
and $S^{\dag}_{\alpha+}=S_{\alpha-}$. $V(\bold r)$ is the external
trap potential. The s-wave scattering potential is characterizes
via $U_{\alpha\beta}(\bold
r)=(4\pi\hbar^2a_{\alpha\beta}/m)\delta^{(3)}(\bold r-\bold r')$
with $a_{\alpha\beta}$ the scattering length.

The interaction Hamiltonian $(H_1+H_2)$ can be diagonalized with a
local unitary transformation. Similar to the former results
\cite{liu}, here we consider the large detuning case i.e.
$\Delta_j^2\gg\Omega_{j0}^2$. In this way, spontaneous emission is
suppressed by introducing the adiabatic condition \cite{sun} that
the population of the higher levels is adiabatically eliminated,
and the total system is restricted to the two ground states
$|S_-\rangle$ and $|S_+\rangle$. Under the present adiabatic
condition the Hamiltonian (\ref{eqn:Hamiltonian1}) can be written
in an effective form which involves only the two ground states:
\begin{eqnarray}\label{eqn:Hamiltonian2}
H&=&\int
d^3r\phi_{s_-}^{*}\bigr[\frac{1}{2m}(i\hbar\partial_k+\frac{e}{c}A_k)^2+V_-(\bold
r)\bigr]
\phi_{s_-}\nonumber\\
&+&\int
d^3r\phi_{s_+}^{*}\bigr[\frac{1}{2m}(i\hbar\partial_k-\frac{e}{c}A_k)^2+V_+(\bold
r)\bigr]
\phi_{s_+}\nonumber\\
&+&\sum_{\mu,\nu=+,-}\int d^3rd^3r'\phi_{s_\mu}^*(\bold
r)\phi_{s_\nu}^*(\bold r')U_{\mu\nu}(\bold r,\bold r')\times\nonumber\\
&&\times\phi_{s_\mu}(\bold r)\phi_{s_\nu}(\bold r').
\end{eqnarray}
Here the vector and scalar potentials induced by the atom-light
couplings are \cite{liu}: $\bold A_-=-\bold A_+=\bold A=\hbar
lce^{-1}\frac{\Omega_0^2}{\Delta^2}(x\hat e_y-y\hat e_x)/\rho^2$
and (neglecting constant terms) $V_\pm(\bold r)=V_{eff}(\bold
r)=V(\bold
r)-\hbar\Omega^2_{0}/\Delta-\hbar^2l^2\Omega_0^4/(2m\Delta^4\rho^2)$
with $\rho=\sqrt{x^2+y^2}$. In the above calculations we have set
$\Delta_1=\Delta_2=\Delta$, $\Omega_{10}=\Omega_{20}=\Omega_0$ and
$l_1=-l_2=l$, i.e. the angular momenta of the two light fields are
opposite in direction. Generally, we assume the total atomic
number is $N=N_++N_-$, where $N_\pm$ are the numbers of atoms in
states $|S_\pm\rangle$. To facilitate further discussion, we
describe here the effective Hamiltonian in the $N$-particle case:
\begin{eqnarray}\label{eqn:Hamiltonian3}
H&=&\sum_{j=1}^{N_+}\bigr[\frac{1}{2m}\bigl(P^{+j}_{k}+\frac{e}{c}A_k(\bold
r^+_{j})\bigl)^2+V_{eff}(\bold
r^+_{j})\bigr]\nonumber\\
&+&\sum_{j=1}^{N_-}\bigr[\frac{1}{2m}\bigl(P^{-j}_{k}-\frac{e}{c}A_k(\bold
r^-_{j})\bigl)^2+V_{eff}(\bold
r^-_{j})\bigr]\nonumber\\
&+&\sum_{j<k}\sum_{\alpha,\beta=+,-}(4\pi\hbar^2a_{\alpha\beta}/m)\delta^{(3)}(\bold
r^\alpha _{j}-\bold r^{\beta}_{k}).
\end{eqnarray}

For convenience, in this paper we shall consider the
spin-independent s-wave scattering, say, $a_{\alpha\beta}=a\equiv
const.$, independent of $\alpha,\beta$. Practically, we apply two
columnar spreading light fields that $\Omega_{01}(\bold
r)=\Omega_{02}(\bold r)=f\rho$ with the coefficient $f>0$. This
kind of fields can be created by e.g. high order Bessel beams
\cite{angular}. Further, we set a two-dimensional harmonic trap by
$V(\bold r)=\frac{1}{2}m\omega^2_\perp\rho^2$, so the scalar
potential reads $V_{eff}(\bold
r)=\frac{1}{2}m\omega^2_{eff}\rho^2$, where
$\omega^2_{eff}=\omega^2_\perp-(1+\frac{\hbar
l^2f^2}{2m\Delta^3})\frac{2\hbar f^2}{m\Delta}$. Note the atomic
numbers in spin-up and spin-down states are determined by initial
condition that can be controlled in experiment. Here we would like
to assume $N_\pm=N/2$. Finally, we can apply a tight harmonic
confinement along $z$-axis with frequency $\omega_z$ such that
$z$-axial ground state energy far exceeds any other transverse
energy scale, yielding a quasi-2D system \cite{LLL}. With these
considerations we can further obtain the effective Hamiltonian by
\begin{eqnarray}\label{eqn:Hamiltonian4}
H&=&\sum_{j=1}^{N/2}\frac{1}{2m}\bigl(\bold P^{+j}+\frac{eB}{2c}
\hat{\bold e}_z\times\bold r^+_{j}\bigl)^2+H_L^+\nonumber\\
&+&\sum_{j=1}^{N/2}\frac{1}{2m}\bigl(\bold P^{-j}-\frac{eB}{2c}
\hat{\bold e}_z\times\bold r^-_{j})\bigl)^2+H_L^-\nonumber\\
&+&\sum_{j<k}\sum_{\alpha,\beta=+,-}\tilde{g}\delta^{(2)}(\bold
r^\alpha _{j}-\bold r^{\beta}_{k}).
\end{eqnarray}
Here $\tilde{g}=a\sqrt{8\pi\hbar^3\omega_z^2/m}$ is the 2D
interaction strength, the angular momentum part reads
\begin{eqnarray}\label{eqn:angularmomentum1}
H_L^{\pm}=\pm(1-\Theta)eBL_z^\pm/4mc
\end{eqnarray}
with the total angular momenta of atoms in spin states
$|S_\pm\rangle$: $L_z^\pm=\sum_{j=1}^{N/2}L_{jz}^{\pm}$ and
$\Theta=(1+\frac{4m^2\Delta^4\omega_{eff}^2}{\hbar^2l^2f^4})^{-1/2}$
equivalent to the ``rotation rate" of fractional quantum Hall
effect (FQHE) in rotating bosonic systems
\cite{Cornell,FQHE1,FQHE2,FQHE} that has been widely studied in
recent years, and
\begin{eqnarray}\label{eqn:field1}
B=\frac{\hbar l
c}{e}\frac{f^2}{\Delta^2}\bigr(1+\frac{4m^2\Delta^2\omega_{eff}^2}{\hbar^2l^2f^4}\bigr)^{\frac{1}{2}}
\end{eqnarray}
characterizes the optically induced magnetic field. From the
formula (\ref{eqn:Hamiltonian4}) one can see the key difference
between our model and FQHE in the rotating BECs
\cite{Cornell,FQHE1,FQHE2,FQHE} is that here atoms experience
$spin$-$dependent$ effective magnetic fields ($\bold B_-=-\bold
B_+=B\hat {\bold e}_z$). In the rotating bosonic atomic system,
even the atomic spin degree is considered, all different spin
states are in the same rotating direction, thus experience only a
single (spin-independent) effective magnetic field. It is also
noteworthy that the charge hall effect system or rotating bosonic
atomic system is $P$-invariant but $T$-breaking. However, our
system is both $P$- and $T$-invariant.

\subsection{Double $\Lambda$-type configuration}
\begin{figure}[ht]
\includegraphics[width=1.0\columnwidth]{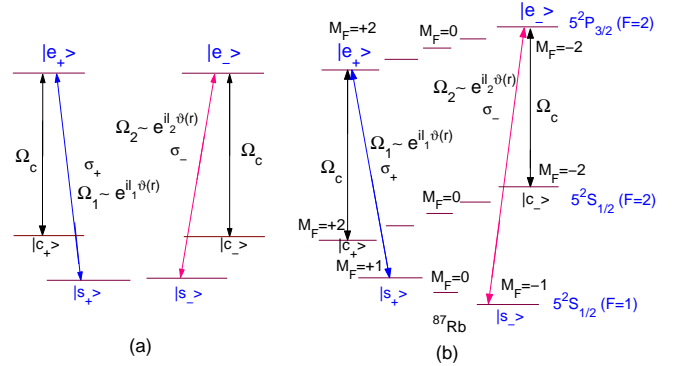}
\caption{(color online) (a) Double $\Lambda$-type bosonic atoms
interacting with two light fields ($\Omega_{1,2}$) with orbital
angular momentum and one strong field $\Omega_c$; (b) This
situation can be experimentally realized with e.g. $^{87}$Rb
atoms. Zeeman splitting is considered.} \label{fig1}\end{figure}
In this subsection we consider another situation, say the double
$\Lambda$-type configuration (see Fig. 2 (a)) to reach the
effective Hamiltonian (\ref{eqn:Hamiltonian4}). The transitions
from $|s_\pm\rangle$ to $|e_\pm\rangle$ are respectively coupled
by a $\sigma_+$ light with the Rabi-frequency
$\Omega_1=\Omega_{10}\exp(i(\bold k_1\cdot\bold r+l_1\vartheta))$
and by a $\sigma_-$ light with the Rabi-frequency
$\Omega_2=\Omega_{20}\exp(i(\bold k_2\cdot\bold r+l_2\vartheta))$,
where $\bold k_{1,2}=k_{1,2}\bold{\hat e}_z$ and
$\vartheta=\tan^{-1}({y}/{x})$. Different from the former
situation, here the couplings are resonant. Besides, we apply the
third strong $\pi$ laser field with
$\Omega_c=\Omega_{c0}\exp(i\bold k_c\cdot\bold r)$ that couples
both transitions from $|c_+\rangle$ to $|e_+\rangle$ and from
$|c_+\rangle$ to $|e_+\rangle$. Also, we introduce the
slowly-varying amplitudes of atomic wave-functions by:
$\phi_{s_\pm}=\Phi_{s_\pm}$, $\phi_{e_\pm}=\Phi_{e_\pm}(\bold
r,t)e^{-i(\bold k_{1,2}\cdot\bold r-\omega_{e_\pm}t)}$,
$\phi_{c_\pm}=\Phi_{c_\pm}(\bold r,t)e^{-i[(\bold k_{1,2}-\bold
k_c)\cdot\bold r-\omega_{c_\pm}]t}$. The total Hamiltonian of the
present system $H=H_0+H_{\Lambda_1}+H_{\Lambda_2}$ is given by
\begin{eqnarray}\label{eqn:Hamiltonian11}
H_0&=&\sum_{\alpha=e_{\pm},s_{\pm},c_\pm}\int
d^3r\phi_{\alpha}^*\bigl(-\frac{\hbar^2}{2m}\nabla^2+V(\bold
r)\bigl)
\phi_{\alpha}\nonumber\\
&+&\sum_{\alpha,\beta}\int d^3rd^3r'\phi_\alpha^*(\bold
r)\phi_\beta^*(\bold r')U_{\alpha\beta}(\bold r,\bold r')
\phi_\alpha(\bold r)\phi_\beta(\bold r'),\nonumber\\
H_{\Lambda_1}&=&-\hbar\int
d^3r(\phi^*_{e_+}\Omega_{10}e^{il_1\vartheta}S_{e_+s_+}\phi_{s_+}+h.a.)\nonumber\\
&&-\hbar\int
d^3r(\phi^*_{e_+}\Omega_{c0}S_{e_+c_+}\phi_{c_+}+h.a.),\\
H_{\Lambda_2}&=&-\hbar\int
d^3r(\phi^*_{e_-}\Omega_{20}e^{il_2\vartheta}S_{e_-s_-}\phi_{s_-}+h.a.)\nonumber\\
&&-\hbar\int
d^3r(\phi^*_{e_-}\Omega_{c0}S_{e_-c_-}\phi_{c_-}+h.a.).\nonumber
\end{eqnarray}
It is easy to check that both $H_{\Lambda_1}$ and $H_{\Lambda_2}$
have three eigenstates, i.e. one dark state and two bright states
\cite{darkstate1,darkstate2}:
$|D_1\rangle=\cos\theta_1|s_+\rangle-\sin\theta_2e^{-il_1\vartheta}|c_+\rangle$,
$|B_{1\pm}\rangle=\bigr[|e_+\rangle\pm(\sin\theta_1|s_+\rangle+\cos\theta_2e^{-il_1\vartheta}|c_+\rangle)\bigr]/\sqrt{2}$
for $H_{\Lambda_1}$ and
$|D_2\rangle=\cos\theta_2|s_+\rangle-\sin\theta_2
e^{-il_2\vartheta}|c_+\rangle$,
$|B_{2\pm}\rangle=\bigr[|e_+\rangle\pm(\sin\theta_2|s_+\rangle+\cos\theta_2
e^{-il_2\vartheta}|c_+\rangle)\bigr]/\sqrt{2}$ for
$H_{\Lambda_2}$, where the mixing angles are defined via
$\tan\theta_{1,2}=|\Omega_{1,2}|/\Omega_{c0}$. The corresponding
eigenvalues are $E_{D_{1,2}}=0$,
$E_{B_{1\pm}}=\pm\sqrt{\Omega_{c0}^2+\Omega_{10}^2}$ and
$E_{B_{2\pm}}=\pm\sqrt{\Omega_{c0}^2+\Omega_{20}^2}$. For our
purpose we require the full system is trapped in the dark-state
subspace $|D_{1,2}\rangle$ (a pseudospin-$1/2$ space), which
excludes the excited states. This can be achieved when the laser
fields are sufficiently strong so that the eigenvalues of the
bright states are far separated from that of the two dark states.
Under this condition the Hamiltonian (\ref{eqn:Hamiltonian11}) can
be written in the effective form which involves only the two dark
states:
\begin{eqnarray}\label{eqn:Hamiltonian12}
H&=&\int
d^3r\phi_{D_1}^{*}\bigr[\frac{1}{2m}(i\hbar\partial_k+\frac{e}{c}A_{1k})^2+V_1(\bold
r)\bigr]
\phi_{D_1}\nonumber\\
&+&\int
d^3r\phi_{D_2}^{*}\bigr[\frac{1}{2m}(i\hbar\partial_k+\frac{e}{c}A_{2k})^2+V_2(\bold
r)\bigr]
\phi_{D_2}\nonumber\\
&+&\sum_{\mu,\nu=1,2}\int d^3rd^3r'\phi_{D_\mu}^*(\bold
r)\phi_{D_\nu}^*(\bold r')U_{\mu\nu}(\bold r,\bold r')\times\nonumber\\
&&\times\phi_{D_\mu}(\bold r)\phi_{D_\nu}(\bold r').
\end{eqnarray}
Here the vector potentials are calculated by $\bold A_{1,2}=i\hbar
c/e\langle D_{1,2}|\nabla|D_{1,2}\rangle=\hbar
l_{1,2}ce^{-1}\sin^2\theta_{1,2}(x\hat e_y-y\hat e_x)/\rho^2$.
Similar as before, we set $l_1=-l_2=l$ and $\Omega_{01}(\bold
r)=\Omega_{02}(\bold r)=f\rho$, while $\Omega_{c0}$ is constant
satisfying $\Omega_{c0}^2\gg|\Omega_{1,2}|^2$. Under this
condition one can find the dark states
$|D_{1,2}\rangle\approx|s_{\pm}\rangle$ and the vector potentials
are followed by $\bold A_{2}=-\bold A_{1}=\bold A=\hbar
f^2lce^{-1}\Omega_{c0}^{-2}(x\hat e_y-y\hat e_x)$. Accordingly,
the scalar potentials are obtained by $V_{1,2}(\bold
r)=V_{eff}(\bold r)\approx V(\bold
r)-\hbar^2l^2f^4/(2m\Omega_{c0}^4)$.

Though a straightforward generalization from the three-level
$\Lambda$ configuration \cite{lambda1,lambda2}, the nontrivialness
of the present double $\Lambda$ bosonic system with spin-dependent
gauge field is protected by the result of quantum SHE whose
integer version is identified to be of $Z_2$ topology
\cite{graphene1,graphene2}. Again, we consider the
spin-independent s-wave scattering, say, $a_{\mu\nu}=a\equiv
const.$, and equal numbers of atoms ($N_1=N_2=N/2$) in the states
$|D_{1,2}\rangle$. When a tight harmonic confinement is applied
along $z$-axis, we can rewrite the above effective Hamiltonian by
\begin{eqnarray}\label{eqn:Hamiltonian14}
H&=&\sum_{j=1}^{N/2}\frac{1}{2m}\bigl(\bold P_{1j}+\frac{eB}{2c}
\hat{\bold e}_z\times\bold r_{1j}\bigl)^2+H_{1L}\nonumber\\
&+&\sum_{j=1}^{N/2}\frac{1}{2m}\bigl(\bold P_{2j}-\frac{eB}{2c}
\hat{\bold e}_z\times\bold r_{2j})\bigl)^2+H_{2L}\nonumber\\
&+&\sum_{j<k}\sum_{\alpha,\beta=1,2}\tilde{g}\delta^{(2)}(\bold
r^\alpha _{j}-\bold r^{\beta}_{k}).
\end{eqnarray}
The parameters in above formula can be similarly obtained as done
in Eqs. (\ref{eqn:angularmomentum1}) and (\ref{eqn:field1}), say
$\tilde{g}=a\sqrt{8\pi\hbar^3\omega_z^2/m}$, the angular momentum
part $H_{1L,2L}=\pm(1-\Theta)eBL_z^\pm/4mc$ with the total angular
momenta of atoms in pseudospin states $|D_{1,2}\rangle$:
$L_z^\pm=\sum_{j=1}^{N/2}L_{jz}^{\pm}$ and
$\Theta=(1+\frac{4m^2\Omega_{c0}^4\omega_{eff}^2}{\hbar^2l^2f^4})^{-1/2}$,
and $B=\frac{\hbar l
c}{e}\frac{f^2}{\Omega_{c0}^2}\bigr(1+\frac{4m^2\Omega_{c0}^2\omega_{eff}^2}{\hbar^2l^2f^4}\bigr)^{\frac{1}{2}}$.
It is clear that the effective Hamiltonian
(\ref{eqn:Hamiltonian14}) is equivalent to that obtained in Eq.
(\ref{eqn:Hamiltonian4}).

\section{FSH state and quasi-particle excitation}

Atoms in different spin states experience the opposite magnetic
fields $\bold B_\alpha$. This leads to a Landau level structure
for each spin orientation. Together with the nonlinear
interactions between spin states, the Hamiltonian
(\ref{eqn:Hamiltonian4}) or (\ref{eqn:Hamiltonian14}) describes a
FSHE in the bosonic system.

\subsection{FSH state}

In this subsection we shall first derive the FSH states for our
system, and then in the next one discuss the related
quasi-particle excitation. For this we consider the large optical
angular momentum condition so that $\omega_{eff}\ll\omega=eB/mc$,
then we approach the limit $\Theta\rightarrow1$, which, in fact,
corresponds to the fast rotating condition in usual bosonic atomic
systems. In this way, the energy scales characterizing Hamiltonian
$H_L^\pm$ are much smaller than those corresponding to other parts
of $H$. Besides, we consider the case that atomic interaction
energy is smaller than the energy spacing between two neighbor
Landau levels. The two restrictions lead to LLL condition in our
system (we shall return to the validity of this approximation
later). The ground state and elementary excitations of
(\ref{eqn:Hamiltonian4}) will then lie on the subspace of common
zero energy eigenstates of $H-H_L^\pm$ \cite{FQHE1,FQHE2}. For
this we can write down the many-body function of the present
system as:
\begin{eqnarray}\label{eqn:function0}
\Psi(z,\varpi^*)&=&{\cal P}(z_1,z_2,...,z_{N/2};\varpi_1^*,\varpi_2^*,...,\varpi_{N/2}^*)\times\nonumber\\
&&\times\prod_{j,k}e^{-|z_j|^2/2-|\varpi_k|^2/2},
\end{eqnarray}
where $z=x^++iy^+$, $\varpi=x^-+iy^-$ are respectively the
coordinates of atoms in states $|S_+\rangle$ and $|S_-\rangle$,
and ${\cal P}(z;\varpi^*)$ is a polynomial in all atomic
coordinates. Denoting by $(z_{ij},Z_{ij})$, $(\varpi_{ij},W_{ij})$
the relative and center mass coordinates of spin-up and spin-down
atoms, and $(t_{ij},T_{ij})$ the relative and center mass
coordinates of the $i$-th spin-up and the $j$-th spin-down atoms,
we can expand the polynomial as
\begin{eqnarray}\label{eqn:function00}
{\cal
P}(z;\varpi^*)=\sum_{m,n,k}f_mg_nh_k\prod_{i<j}^{N/2}z_{ij}^m\prod_{k<l}^{N/2}
\varpi_{kl}^{*n}\prod_{u,v}^{N/2}t_{\mu\nu}^w.
\end{eqnarray}
To ensure the function $\Psi(z,\varpi^*)$ is a zero-eigenstate of
the nonlinear interaction Hamiltonian, $f_0, g_0$ and $h_0$ must
be zero. Furthermore, the interchange symmetry of bosonic atoms
determines that $m, n$ must be even integers. Then, $z_{ij}^2$ and
$\varpi_{ij}^{*2}$ are the factors of (\ref{eqn:function00}), and
we can rewrite the many-body wave function by
\begin{eqnarray}\label{eqn:function1}
\Psi(z,\varpi^*)&=&Q(z,\varpi^*)\prod_{i<j}^{N/2}(z_i-z_j)^2\prod_{k<l}^{N/2}(\varpi^*_k-\varpi^*_l)^2\times\nonumber\\
&\times&\prod_{u,v}^{N/2}(z_u-\varpi_v^*)^w\prod_{j,k}e^{-|z_j|^2/2-|\varpi_k|^2/2},
\end{eqnarray}
where $w\geq1$ is an arbitrary positive integer. The similar state
has been studied in the electronic system \cite{zhang2}. The
formula (\ref{eqn:function1}) can be divided into two basic types
of many-body functions depending on $w$ takes odd integers (1st
type) and even integers (2nd type) respectively. It is easy to see
the 1st type of function is antisymmetric upon the interchange
$z\longleftrightarrow\varpi^*$ reflecting the $|S_+\rangle$ chiral
- $|S_-\rangle$ chiral $antisymmetry$, while the 2nd type is
symmetric upon the interchange $z\longleftrightarrow\varpi^*$ that
 reflects the $|S_+\rangle$ chiral
- $|S_-\rangle$ chiral $symmetry$. $H_L^\pm$ can be diagonalized
within the truncated Hilbert space specified by
(\ref{eqn:function0}) and (\ref{eqn:function1}). In our model, it
is interesting that when $Q(z;\varpi^*)$ is a homogeneous
polynomial in $z_{ij}, \varpi_{ij}^*$ and $t_{ij}$, the wave
function $\Psi(z,\varpi^*)$ is an eigenstate of $H_L^++H_L^-$ with
the eigenvalue
\begin{eqnarray}\label{eqn:energy}
E_L=(1-\Theta)eB(M_+-M_-)/4mc.
\end{eqnarray}
Note $M_+>0$ (in the $+z$ direction) and $M_-<0$ (in the $-z$
direction) are respectively total angular momenta of spin-up and
spin-down atoms. Therefore, the ground state of our system is
determined by the angular momentum difference between spin-up and
spin-down atoms, say, for the 1st type, the ground state
corresponds to $Q(z,\varpi^*)=1$ and $w=1$ so that
\begin{eqnarray}\label{eqn:function2}
\Psi^{(1)}(z,\varpi^*)&=&\prod_{i<j}^{N/2}(z_i-z_j)^2\prod_{k<l}^{N/2}(\varpi^*_k-\varpi^*_l)^2\times\nonumber\\
&\times&\prod_{u,v}^{N/2}(z_u-\varpi_v^*)\prod_{j,k}e^{-|z_j|^2/2-|\varpi_k|^2/2}.
\end{eqnarray}

The 1st type of ground state (\ref{eqn:function2}) of the present
system has several fundamental properties. Firstly, this state is
analogous to the Halperin's function of two different spin states
\cite{halperin}, but here the two spins experience opposite
effective magnetic fields. Secondly, the filling factor in the
FSHE is defined by the ratio between the total number of atoms
($N$) and the number of total orbital angular momentum states
($M_+-M_-$). One can verify that the filling factor of our system
is given by
\begin{eqnarray}\label{eqn:factor1}
\bar\nu=\frac{2}{m+n+2w},
\end{eqnarray}
where, according to the Eq. (\ref{eqn:function2}), $m=2,n=2$ and
$w=1$. Thus the filling factor of the 1st type of ground state
$\bar\nu=1/3$. This factor is well-known in the quantum hall
electron system \cite{electron}. However, for bosonic system, this
result may lead to new insights. Thirdly, noting that
(\ref{eqn:function2}) is the spatial wave function, the total wave
function is obtained by multiplying it by the spin part
\begin{eqnarray}\label{eqn:spinpart1}
{\cal
A}(\chi)&=&\sum^N_{i_1...i_{N/2};j_1...j_{N/2}}\prod_{i_\alpha<i_{\alpha'}}^{i_{N/2}}(1-\delta_{i_\alpha
i_{\alpha'}})\prod_{j_\beta<j_{\beta'}}^{j_{N/2}}(1-\delta_{j_\beta
j_{\beta'}})\nonumber\\
&\times&\prod_{i_\alpha=i_1,j_\beta=j_1}^{i_{N/2},j_{N/2}}\epsilon_{i_\alpha
j_\beta}\chi_{i_\alpha}^+\chi_{j_\beta}^-,
\end{eqnarray}
which has also $|S_+\rangle$ chiral - $|S_-\rangle$ chiral
$antisymmetry$. Here $\chi^{\pm}$ are the spinor components of
atoms and $\epsilon_{i_\alpha j_\beta}$ equals $+1$ for
$i_\alpha<j_\beta$, $-1$ for $i_\alpha>j_\beta$ and $0$ for
$i_\alpha=j_\beta$. Finally, the angular momentum of spin-up or
spin-down atoms or their total angular momentum is not conserved.
Nevertheless, it is interesting that their angular momentum
difference $(L_z^+-L_z^-)$ is conserved. One can verify the
relation $L_z^+-L_z^-=N(N-1)/(2\bar\nu)$ for our system.

Furthermore, we discuss the many-body function of the 2nd type.
Similar to the previous discussion, one can show the 2nd type of
ground state corresponds to $Q(z,\varpi^*)=1$ and $w=2$. Thus we
have
\begin{eqnarray}\label{eqn:function3}
\Psi^{(2)}(z,\varpi^*)&=&\prod_{i<j}^{N/2}(z_i-z_j)^2\prod_{k<l}^{N/2}(\varpi^*_k-\varpi^*_l)^2\times\nonumber\\
&\times&\prod_{u,v}^{N/2}(z_u-\varpi_v^*)^2\prod_{j,k}e^{-|z_j|^2/2-|\varpi_k|^2/2}.
\end{eqnarray}
Different from $\Psi^{(1)}$, this state has the property of
$|S_+\rangle$ and $|S_-\rangle$ chiral $symmetry$. The total wave
function of the 2nd type can be obtained by multiplying it by the
spin part
\begin{eqnarray}\label{eqn:spinpart2}
{\cal S}(\chi)&=&\sum^N_{i_1...i_{N/2};j_1...j_{N/2}}
\prod_{i_\alpha<i_{\alpha'}<j_\beta<j_{\beta'}}^{i_{N/2},j_{N/2}}(1-\delta_{i_\alpha
i_{\alpha'}})\nonumber\\
&\times&(1-\delta_{j_\beta
j_{\beta'}})\prod_{i_\alpha<j_\beta}^{i_{N/2},j_{N/2}}\chi_{i_\alpha}^+\chi_{j_\beta}^-,
\end{eqnarray}
which has $|S_+\rangle$ chiral - $|S_-\rangle$ chiral $symmetry$.
The filling factor of this state is easy to obtained by setting
$m=n=w=2$ in the Eq. (\ref{eqn:factor1}), so we get $\bar\nu=1/4$.
It is easy to see that the energy of $\Psi^{(1)}$ is smaller than
$\Psi^{(2)}$. However, the optical transition between any two
states of different types is forbidden due to the different chiral
symmetries. Therefore, both type of ground states can be
adiabatically stable.

Before ending this subsection, we point out that when an effective
in-plane electric field is applied through, e.g. optical means
\cite{liu} or through the gravity \cite{zhu}, we shall obtain a
transverse spin current. Since the center-of-mass motion is
independent of the atom-atom interaction, the SHC is solely
determined by the filling factors, similar to the charge Hall
conductivity in the fractional quantum Hall effect (FQHE)
\cite{FQHE3}. For this we have FSH conductivity
$\sigma^{SH}_{xy}=2\nu\frac{e}{2\pi}$. Here we keep the factor
``$2$" to indicate the FSH conductivity is contributed from both
spin-up and spin-down species and is then doubled \cite{zhang2}.
On the other hand, the charge Hall conductivity is always zero due
to the time-reversal symmetry of the system.

\subsection{Quasi-particle excitation and fractional statistics}

The FSH state obtained above can be detected by measuring the
fractional statistical phase of the quasi-particles with a
Ramsey-type interferometer proposed in \cite{FQHE1}. The
quasi-hole excitation can be obtained by inserting a laser in the
system that create localized repulsive potential, analogy to an
impurity with positive $\delta$-potential, in the atomic gas.
Specifically, if we consecutively apply such two lasers
respectively at position $\eta_0$ and $\eta_1$, we can
adiabatically evolve the initial ground many-body state, say
$\Psi^{(1)}(z,\varpi^*)$ of the first type to the superposition of
the one- and two-quasi-hole state
$\Psi^{(1)}\sim\Psi^{(1)}_{\eta_0}+\Psi^{(1)}_{\eta_0,\eta_1}$.
Then we adiabatically move the laser initially at position
$\eta_0$ along a closed path enclosing position $\eta_1$, and at
the end of the process we get the final state by \cite{FQHE1}
\begin{eqnarray}\label{eqn:function4}
\Psi^{(1)}_F(z,\varpi^*)\sim\Psi^{(1)}_{\eta_0}+e^{i\gamma^{(1)}}\Psi^{(1)}_{\eta_0,\eta_1},
\end{eqnarray}
where
$\gamma=i\oint_C\langle\Psi^{(1)}_F|\partial_{\eta_0}|\Psi^{(1)}_F\rangle
d\eta_0$ is the statistical phase characterizing the quasi-holes.
For the present FSH regime, this phase has three different results
depending on the types of the created quasi-holes. Firstly, if the
lasers at position $\eta_0$ and $\eta_1$ couple only to the
$spin$-$up$ atoms, the quasi-particles at $\eta_0$ and $\eta_1$
correspond to spin-up atoms, say
\begin{eqnarray}\label{eqn:quasifunction4}
\Psi^{(1)}_{\eta_0}&=&\prod_{j}^{N/2}(z_j-\eta_0)\Psi^{(1)}(z,\varpi^*),\nonumber\\
\Psi^{(1)}_{\eta_0,\eta_1}&=&\prod_{j,k}^{N/2}(z_j-\eta_0)(z_k-\eta_1)\Psi^{(1)}(z,\varpi^*),
\end{eqnarray}
and we obtain the statistical phase $\gamma_1^{(1)}=2\pi/3$. The
interchange of such two quasi-holes then gives the fractional
phase $\pi/3$, which identifies the $1/3$-anyon. Secondly, if the
lasers at position $\eta_0$ and $\eta_1$ couple only to the
$spin$-$down$ atoms, the quasi-holes are obtained by a simple
transformation $z\longleftrightarrow\varpi^*$ in the Eq.
(\ref{eqn:quasifunction4}) (i.e.
$\Psi^{(1)}_{\eta_0}=\prod_{u}^{N/2}(\varpi^*_u-\eta_0)\Psi^{(1)},
\Psi^{(1)}_{\eta_0,\eta_1}=\prod_{u,v}^{N/2}(\varpi^*_u-\eta_0)(\varpi^*_v-\eta_1)\Psi^{(1)}$),
and we shall obtain the statistical phase
$\gamma_2^{(1)}=-2\pi/3$, which is equivalent to $4\pi/3$ and
reflects the spin-down atoms experience the effective magnetic
field opposite to that the spin-up atoms do. Finally, if the two
lasers couple to both the $spin$-$up$ and $spin$-$down$ atoms, we
have
\begin{eqnarray}\label{eqn:quasifunction5}
\Psi^{(1)}_{\eta_0}&=&\prod_{j,u}^{N/2}(z_j-\eta_0)(\varpi^*_u-\eta_0)\Psi^{(1)}(z,\varpi^*),\nonumber\\
\Psi^{(1)}_{\eta_0,\eta_1}&=&\prod_{j,k}^{N/2}(z_j-\eta_0)(z_k-\eta_1)\times\nonumber\\
&&\times\prod_{u,v}^{N/2}(\varpi^*_u-\eta_0)(\varpi^*_v-\eta_1)\Psi^{(1)}(z,\varpi^*),
\end{eqnarray}
and the statistical phase for the quasi-holes can be calculated as
$\gamma_3^{(1)}=0$. In this case the quasi-particle becomes boson.
Accordingly, the statistical phases for the second type ground
state are obtained by $\gamma_1^{(2)}=-\gamma_2^{(2)}=\pi/2$ and
$\gamma_3^{(2)}=0$. The zero phase in the third case actually
explains the charge Hall conductivity should be zero in the
quantum SHE. The statistical phase can be detected via a
Ramsey-type interferometer. As a comparison, in the FQH regime,
the first type ground state has the filling factor $\nu=2/3$ and
in the above process one can obtain the statistical phases
$\gamma_1=\gamma_2=2\pi/3$ and $\gamma_3=4\pi/3$ in the three
different cases \cite{FQHE1}. As a result, the present FSHE can be
distinguished from FQH regime in the measurement.

Now we discuss the restrictions of LLL condition employed in our
system. The validity of LLL approximation used in previous
discussions is determined by three considerations. Firstly, the
energy corresponding to angular momentum, $\epsilon_l=E_L/N$
should be smaller than the interaction energy per particle,
$\epsilon_{int}=\beta n_a\tilde{g}$, where the coefficient
$\beta\approx1$ and
$n_a=\langle|\phi_{s_+}|^2+|\phi_{s_-}|^2\rangle\approx\sqrt{mN\omega^2(1-\Theta)/\tilde{g}}$
is the atomic average density \cite{LLL}. Furthermore, the later
energy should also be smaller than the spacing between Landau
levels $\epsilon_{lan}=\hbar\omega$. It then follows from the two
requirements that
\begin{eqnarray}\label{eqn:inequality}
N\ll min\{\frac{\Theta^2}{1-\Theta^2}\frac{\hbar^2}{\tilde{g}m}, \
\frac{64\bar\nu^2}{1-\Theta^2}\frac{\tilde{g}m}{\hbar^2}\}.
\end{eqnarray}
For weakly interacting case ($\tilde{g}\ll\hbar^2/m$), this
inequality reads
$N\ll\frac{64\bar\nu^2}{1-\Theta^2}\frac{\tilde{g}m}{\hbar^2}$,
and for strongly interacting case ($\tilde{g}\geq\hbar^2/m$), one
has $N\ll\frac{\Theta^2}{1-\Theta^2}\frac{\hbar^2}{\tilde{g}m}$.
Besides, another condition is that the effective magnetic flux
induced by light fields can support a sufficiently large number of
vortices, for which the boundary effect of the system can be
neglected.

\section{Experimental conditions}

First, we discuss the experimental realization of the four-level
and double $\Lambda$-type systems discussed above. Candidate atoms
include $^{87}$Rb, $^{23}$Na and $^7$Li bosonic systems. As an
example, we consider first the $^{87}$Rb atomic system. For the
four-level system in Fig. 1(b) we employ the transitions
$(5^2S_{1/2},F=1)\rightarrow(5^2P_{1/2},F=1)$ and
$(5^2S_{1/2},F=1)\rightarrow(5^2P_{3/2},F=1)$. The two ground
states $|S_+\rangle$ and $|S_-\rangle$ correspond to
$|F=1,M_F=+1\rangle$ and $|F=1,M_F=-1\rangle$, respectively, while
both $|e_+\rangle$ and $|e_-\rangle$ correspond to
$|F=1,M_F=0\rangle$ chosen from $5^2P_{1/2}$ and $5^2P_{3/2}$.
Note another ground sub-level $|F=1,M_F=0\rangle$ can also be
coupled to the excited sub-level $|F=1,M_F=+1\rangle \
(5^2P_{3/2})$ by the $\sigma_+$ light, and to the
$|F=1,M_F=-1\rangle \ (5^2P_{1/2})$ by the $\sigma_-$ light (Fig.
1(c)). However, one can verify that the induced effective gauge
potential on this state ($M_F=0, \ 5^2S_{1/2})$ is proportional to
$\Omega_{10}^2l_1/\Delta_1^2+\Omega_{20}^2l_2/\Delta_2^2=0$.
Furthermore, through the optically stimulated Raman passage in the
$\Lambda$-type configuration in Fig. 1(d), initially one can pump
all atoms into the sub-levels $(M_F=\pm1)$ with equal atomic
numbers by setting $|\Omega_+|=|\Omega_-|$, while population of
$(M_F=0)$ is negligible \cite{darkstate1}. Based on these results,
we can safely neglect the effects of the sub-level $(M_F=0)$ and
only include $|S_\pm\rangle$ $(M_F=\pm1)$ in our system.
Simultaneously turning off $\Omega_\pm$ and then employing the
far-detuning angular-momentum light fields $\Omega_{1,2}$ to the
system, one can reach the Hamiltonian (\ref{eqn:Hamiltonian1}) and
then (\ref{eqn:Hamiltonian3}-\ref{eqn:Hamiltonian4}) for our
model.

The double $\Lambda$-type system can also be realized with
$^{87}$Rb (or $^{23}$Na) atoms, which is shown in Fig. 2(b).
Although principally the state $|F=1,M_F=0\rangle$ ($5^2S_{1/2}$)
can be coupled by the laser fields $\Omega_1$ and $\Omega_2$, the
induced gauge field for this state, similar as the above result,
is also zero. Furthermore, initially one can also optically pump
all atoms into the sub-levels $(M_F=\pm1)$, not into the state
$(M_F=0)$, as done in the four-level configuration. In this way,
the effects of the sub-level $(M_F=0)$ can still be neglected in
the double $\Lambda$ system.

Finally, we turn to the numerical estimate of our results. Again
we consider first the four-level configuration. The energy
splitting between $5^2P_{3/2} (F=1)$ and $5^2P_{1/2} (F=1)$
($\Delta E_1=4.5\times10^4$ GHz) is much larger than that between
$5^2S_{1/2} (F=2)$ and $5^2S_{1/2} (F=1)$ ($\Delta
E_2=6.9\times10^3$ MHz) (see Fig. 1(b)). To avoid the couplings
between the state $5^2S_{1/2} (F=2)$ and the excited ones, we need
the Rabi-frequency $\Omega_0$ of the optical fields to be also
much smaller than energy splitting ($\Delta E_2$) between
$5^2S_{1/2} (F=2)$ and $5^2S_{1/2} (F=1)$. Practically, we can
choose $\Delta_{1,2}=0.5$GHz. Besides, other typical values are
taken as $l_1=-l_2\sim10^3$ \cite{angular},
$f\sim2.5$MHz$\cdot\mu$m$^{-1}$. When the spatial scale of the
interaction region is $R\sim2.0\mu m$, the optical
Rabi-frequencies satisfy $\Omega_0^2\ll\Delta^2,(\Delta E_2)^2$.
The cyclone frequency can then be evaluated by $\omega\sim30$Hz.
For the $^{23}$Na system, under the same parameter choice, we
obtain the cyclone frequency $\omega\sim110$Hz. If $\omega_{eff}$
is tuned to be several hertz, we then have $1-\Theta\sim10^{-3}$.
From the inequality (\ref{eqn:inequality}), this numerical result
implies that for the strongly interacting boson atomic gas
$(\tilde{g}\sim\hbar^2/m)$, the number of atoms can be as large as
$10^2$ without violating the LLL condition, and for the weakly
interacting case $(\tilde{g}\sim0.1\hbar^2/m)$ this number is
about ten.

For the double $\Lambda$-type situation, we can set the parameters
that $\Omega_{c0}=1.0\times10^2$MHz, $l_1=-l_2\sim10^3$,
$f\sim1.0$MHz$\cdot\mu$m$^{-1}$. When the spatial scale of the
interaction region is $R\sim10\mu m$, the optical Rabi-frequencies
satisfy $|\Omega_{1,2}|^2\ll\Omega_{c0}^2$. The cyclone frequency
can then be evaluated by $\omega\sim120$Hz for the $^{87}$Rb atoms
and $\omega\sim400$Hz for the $^{23}$Na system. Therefore
$1-\Theta\sim10^{-3}$ when $\omega_{eff}$ is tuned to be the order
of ten hertz. In this case, without violating the LLL condition,
the number of atoms can be as large as $10^{2\sim3}$ for the
strongly interacting boson atomic gas $(\tilde{g}\sim\hbar^2/m)$,
and be a few tens for the weakly interacting case
$(\tilde{g}\sim0.1\hbar^2/m)$. We therefore expect the many-body
functions such as (\ref{eqn:function2}) and (\ref{eqn:function3})
obtained here can be reached with a small number of cold atoms.
Note the adiabatic condition is assumed in our system. Atomic
motion may lead to the transition between the ground eigenstates
and excited ones, which results in decay of the ground states. The
transition rate can be evaluated by \cite{liu,zhu,lambda2}
$\tau\sim|\bold
v\cdot\nabla(\frac{|\Omega_{1,2}|}{\Delta})+l_{1,2}\frac{|\Omega_{1,2}|}{\Delta}\bold
v\cdot\nabla\vartheta(\bold r)|$ for four-level system and
$\tau\sim|\bold
v\cdot\nabla(\frac{|\Omega_{1,2}|}{\Omega_{c0}})+l_{1,2}\frac{|\Omega_{1,2}|}{\Omega_{c0}}\bold
v\cdot\nabla\vartheta(\bold r)|$ for double $\Lambda$ system,
where $\bold v$ is the velocity of the atoms. This transition
leads to the effective decays
$\gamma_{eff}\sim\tau^2\gamma_e/\Delta^2$ and
$\gamma_{eff}\sim\tau^2\gamma_e/\Omega_{c0}^2$ for the four-level
and double $\Lambda$ systems, respectively, with $\gamma_e$ the
decay of the excited states. Typical values of the parameters for
a BEC can be $|\bold v|\sim1.0$cm$\cdot$s$^{-1}$ and
$\gamma_{e}\sim10^{7}$s$^{-1}$. We can then estimate the life time
of the atoms as $\mathcal{T}_D\sim\gamma_{eff}^{-1}\sim1.0$s for
the present systems.

\section{Conclusion}

In conclusion we have proposed the fractional spin hall effect
(FSHE) in neutral atomic system by coupling the atomic spin states
(internal angular momentum states) to optical fields. We studied
fundamental properties of the many-body wave function of the
present system under the LLL condition. Especially, we show two
different types of ground states in our system. The first type of
ground state is a $1/3$-factor Laughlin function, and exhibits
chiral-anti-chiral interchange antisymmetry, while the second type
of ground state is a $1/4$-factor wave function with
chiral-anti-chiral symmetry. The fractional statistics of
quasi-particles in the present FSH state are studied, and are
discovered to be different from that of the corresponding FQH
state. Thus the present FSHE can be distinguished from FQH regime
in the measurement. Realization of the present model in realistic
atomic systems was also studied.

\section{Acknowledgements}

We thank Prof Shou-Cheng Zhang for stimulating discussions, and
for his careful reading of the first draft of this work. X.J. Liu
and X. Liu also thank Prof Jairo Sinova and Prof Chia-Ren Hu for
helpful communications about this topic. This work is supported by
NUS academic research Grant No. WBS: R-144-000-189-305, by US NSF
Grant No. DMR-0547875, and by ONR under Grant No.
ONR-N000140610122.

%%%%%%%%%%%%%%%%%%%%%%%%%%%%%%%%%%%%%%%%%%%%%

%%%%%%%%%%%%%%%%%%%%%%%%%%%%%%%%%%%%%%%%%%%%%
\noindent\\

\end{document}